\documentclass[12pt,onecolumn]{IEEEtran}
\usepackage{epsfig, multirow, array, cite}
\usepackage{amssymb, amsmath}
\usepackage{epstopdf}

\ifCLASSINFOpdf
\else
\fi
\begin{document}
\newcommand\comb[2]{\ensuremath\vphantom{\mathrm C}_{#1}
        \kern -.1em\mathrm{C}\kern -.09em{}_{#2}}

\title{Ergodic Interference
Alignment with Delayed Feedback}
\author{\IEEEauthorblockN{Myung Gil Kang, \textit{Student Member, IEEE}, and Wan Choi, \textit{Senior Member, IEEE}}
\thanks{M. G. Kang and W.~Choi are with Department of Electrical
    Engineering, Korea Advanced Institute of Science and Technology
    (KAIST), Daejeon 305-701, Korea (e-mail: casutar@kaist.ac.kr,
    wchoi@ee.kaist.ac.kr).}
}

\maketitle
\newtheorem{lemma}{Lemma}
\newtheorem{theorem}{Theorem}
\newtheorem{definition}{Definition}
\newtheorem{corollary}{Corollary}
\newtheorem{remark}{Remark}
\begin{abstract}
We propose new ergodic interference alignment techniques for
$K$-user interference channels with delayed feedback. Two delayed
feedback scenarios are considered -- delayed channel information at
transmitter (CIT) and delayed output feedback. It is proved that the
proposed techniques achieve total $2K/(K+2)$ DoF which is higher
than that by the retrospective interference alignment for for the
delayed feedback scenarios.

\end{abstract}

\section{Introduction}
In these days, interference management is one of the most important
issues in wireless communication systems. In order to obtain high
spectral efficiency, many interference management techniques have
been proposed and studied. For the two-user interference channel,
the capacity region is already known for weak and strong
interference regions in \cite{lowic} and \cite{costa}. 
For the moderate region, the capacity
region is still unknown, but there are some works that the capacity
regions can be achieved by rate-splitting within one bit
\cite{Tse}. The authors in \cite{Tse} also proved that the optimal
generalized degrees of freedom are achievable using the
rate-splitting scheme.

Compared to the two-user interference channel, the general $K$-user
interference channel have not been much known yet. Many researchers
have studied degrees of freedom (DoF) to understand the asymptotic
capacity because of the difficulty of finding the exact capacity
region. For the $K$-user interference channel, the DoF was shown to
be upper bounded by $\frac{K}{2}$  in \cite{Host}. The authors in
\cite{Jafar} showed that this upper bound can be achieved by the
interference alignment (IA) scheme that all interfering signals from
other transmitters are aligned in the same dimension to
independently decode the desired signals at the receivers. This
scheme operates in high SNR to guarantee independence between the
desired signal dimension and the interference aligned dimension. In
order to operate in any finite SNR, \cite{EIA} proposed ergodic
IA that all interfering signals are perfectly
cancelled out by properly choosing two time indices. Using ergodic
IA, each user can achieve half the interference-free ergodic
capacity.

The IA schemes generally require perfect channel state information
(CSI). In rapidly time varying channels, however, channel state
information becomes outdated due to feedback delay. In other words,
it is impractical to assume that transmitters have perfect knowledge
of current channel state information. In order to solve this
problem, recent studies \cite{HJSV,BIA,RIA,RIAK} focus on
exploiting imperfect channel state information -- no channel state
information at transmitter (CSIT) or delayed feedback information.
It was shown in \cite{RIA} that $9/8$ DoF is achievable for the
three-user interference channel with delayed CSIT and $6/5$ DoF is
achievable for the three-user interference channel with delayed
output feedback without CSIT. More generally, \cite{RIAK} showed
that $K^2/(K^2-1)$ DoF is achievable for the $K$-user interference
channel $(K\ge3)$ with delayed CSIT and $\lceil K/2\rceil K/(\lceil
K/2\rceil(K-1)+1)$ DoF is achievable for the $K$-user interference
channel with delayed output feedback without CSIT.

In this paper, we assume two delayed feedback scenarios as follows.
(i) \emph{Delayed channel information at transmitter}: in this
scenario, nothing but the past channel information is given at the
transmitter. The channel information implies either channel state
information or time index information. Output feedback is not
assumed in this case. (ii) \emph{Delayed output feedback without
CSIT}: in this scenario, nothing but the past output feedback
information is given at the transmitter and the channel information
is not available at the transmitter. We devise effective
interference management strategies in the $K$-user interference
channel for these two scenarios. The proposed schemes are developed
in the framework of ergodic IA \cite{EIA} and enables
interference-free decoding of the desired message at the receiver.
It is shown that the proposed strategies achieve $\frac{2K}{K+2}$
DoF in the $K$-user interference channel for the scenarios of the
delayed channel information and the delayed output feedback without
CSIT. The proposed schemes achieve higher DoF than the retrospective
IA \cite{RIA,RIAK} in the $K$ user interference channel with the
same assumptions of delayed feedback.


\section{System model and preliminary}
\subsection{Interference channel model}
The received signal in the $K$-user interference channel is given by
\begin{align}
\mathbf{Y}(t) = \mathbf{H}(t)\mathbf{X}(t)+\mathbf{Z}(t)
\end{align}
where $\mathbf{Y}(t) = [Y_1(t)~Y_2(t)~\cdots~Y_K(t)]^T$, the
transmitted signal vector $\textbf{X}(t) =  [X_1(t)~X_2(t)$
$~\cdots~X_K(t)]^T\in  \mathbb{C}^{K \times 1} $ with power
constraint $P$, $\mathbf{H}(t) \in \mathbb{C}^{K \times K}$
represents the time varying fading channel matrix and is given by
\begin{align}
\mathbf{H}(t) = \left[
                  \begin{array}{ccc}
                    H_{11}(t) & \cdots & H_{1K}(t) \\
                    \vdots & \ddots & \vdots~ \\
                    H_{K1}(t) & \cdots & H_{KK}(t) \\
                  \end{array}
                \right].
\end{align}
where $H_{ji}$ denotes the channel coefficient from transmitter $i$
to receiver $j$ and is an independent and identically distributed
(i.i.d.) complex Gaussian random variable with distribution $\sim
\mathcal{CN}(0,1)$. At receivers, full channel state information is
assumed to be available, i.e., CSIR. The element of the additive
white Gaussian noise vector $\textbf{Z}(t) =
[Z_1(t)~Z_2(t)~\cdots~Z_K(t)]^T$ is assumed to follow complex
Gaussian distribution $\sim\mathcal{CN}(0,N_0)$.

\subsection{Preliminary: Ergodic IA with Full CSIT}
The ergodic IA \cite{EIA} requires perfect knowledge of channel
state information at the transmitter (CSIT). For a $K$-user
interference channel, $K/2$ total DoF can be achieved in an ergodic
sense if the channel is time varying. Contrary to other IA
techniques, the ergodic IA works for infinite SNR as well as any
finite SNR since interfering signals are canceled out when the
channel matrices at two different time instants satisfy a certain
condition. Specifically, let $t_1$ and $t_2$ be the time instants
(or time indices) at which the the channel matrices satisfy the
following relationship:
\begin{align}
\textbf{H}(t_1) &= \left[
                  \begin{array}{ccc}
                    H_{11}(t_1) & \cdots & H_{1K}(t_1) \\
                    \vdots & \ddots & \vdots~ \\
                    H_{K1}(t_1) & \cdots & H_{KK}(t_1) \\
                  \end{array}
                \right]\label{eq:channel1}\\
                \textbf{H}(t_2) &= c(t_2)\cdot\left[
                  \begin{array}{ccc}
                    H_{11}(t_1) & \cdots & -H_{1K}(t_1) \\
                    \vdots & \ddots & \vdots~ \\
                    -H_{K1}(t_1) & \cdots & H_{KK}(t_1) \\
                  \end{array}
                \right] \label{eq:channel2}
\end{align} where $c(t_2)$ is a complex valued constant and $H_{kk}(t_2) = c(t_2)H_{kk}(t_1)$, $H_{kj}(t_2) = -c(t_2)H_{kj}(t_1)$, $k\neq j$, $k,j\in{1,\ldots,K}$.
At the time $t_2$, the message which was previously sent at the time
$t_1$ is again sent from the transmitter. In other words, the
transmitted signal vector $\mathbf{X}(t_1)$ is equal to
$\mathbf{X}(t_2)$. To decode the message, receiver $k$ adds the
received signals at $t_1$ and $t_2$ and constructs a sufficient
statistics for the message $X_k(t_1)$ as
\begin{align}
Y_k(t_1)+Y_k(t_2)/c(t_2) =
2H_{kk}(t_1)X_k(t_1)+Z_k(t_1)+Z_k(t_2)/c(t_2).
\end{align} Then, the achievable rate is determined by
\begin{align}
R_k = \frac{1}{2}\log(1+\frac{2|H_{kk}|^2}{(1+1/c(t_2)^2)}SNR)-\epsilon
\end{align}
where $SNR = \frac{P}{N_0}$, $\epsilon > 0$. Correspondingly, the
total $\frac{K}{2}$ DoF is achievable \cite{EIA}.

\section{Ergodic interference alignment with delayed CIT}
Contrary to the existing ergodic IA, we assume only full CSIR and
imperfect or partial CSIT by feedback delay. Specifically, all
receivers feed either channel state information or time indices back
to the transmitters. Each transmitter cannot use the current channel
information but can use the outdated channel information due to
feedback delay.

\subsection{Three-user interference channel with delayed CSIT}
In this subsection, we propose a new strategy to achieve high DoF
when the receivers feed CSI back to the transmitters. To effectively
establish the concept of the proposed scheme, we start from the
three-user interference channel, i.e., $K =3$.
\begin{theorem}
Three-user interference channel with delayed CSIT can achieve total
$\frac{6}{5}$ DoF by ergodic IA.
\end{theorem}
\begin{IEEEproof}
The proposed ergodic IA is carried out over two phases:

\textit{Transmission phase 1}: Transmission phase 1 for data
transmission is continued until $t_2$. At each time until $t_2$, new
messages are continuously transmitted. For the time $t_1$ and $ t_2$
at which the channel condition in (\ref{eq:channel1}) and
(\ref{eq:channel2}) is satisfied, the received signals are given by
\begin{align}
\mathbf{Y}(t_1)=\mathbf{H}(t_1)\mathbf{X}(t_1)+\mathbf{Z}(t_1),\label{rs1}\\
\mathbf{Y}(t_2)=\mathbf{H}(t_2)\mathbf{X}(t_2)+\mathbf{Z}(t_2).\label{rs2}
\end{align}
Due to delayed CSIT, the transmitters cannot recognize what the
current channel states are so that they cannot send the same message
of $t_1$ at $t_2$ as in the conventional ergodic IA. Thus, they just
send independent messages at $t_1$ and $t_2$. After the channel
changes, the transmitters can figure out that the channel condition
in (\ref{eq:channel1}) and (\ref{eq:channel2}) is satisfied at the
previous time $t_2$ due to delayed CSI feedback. Then, transmission
phase 2 is entered.

\textit{Transmission phase 2}: If the previous time was $t_2$, the
transmitters send the following signals at time $t_2+1,t_2+2,t_2+3$,
respectively:
\begin{align}
&\mathrm{Transmitter~1~at~time}~t_2+1:~X_1(t_1)-X_1(t_2)\nonumber\\
&\mathrm{Transmitter~2~at~time}~t_2+2:~X_2(t_1)-X_2(t_2)\nonumber\\
&\mathrm{Transmitter~3~at~time}~t_2+3:~X_3(t_1)-X_3(t_2).\nonumber
\end{align}
After transmission phase 2 is completed, the transmission mode goes
back to transmission phase 1. Each receiver adds its received
signals at $t_1$ and $t_2$ and constructs the following variables.
\begin{align}
Y_1(t_1)+Y_1(t_2)/c(t_2) &=  H_{11}(t_1)(X_1(t_1)+X_1(t_2))+H_{12}(t_1)(X_2(t_1)-X_2(t_2))+H_{13}(t_1)(X_3(t_1)-X_3(t_2))\nonumber\\
&~+Z_1(t_1)+Z_1(t_2)/c(t_2),\label{eq:rx1}\\
Y_2(t_1)+Y_2(t_2)/c(t_2) &=  H_{22}(t_1)(X_2(t_1)+X_2(t_2))+H_{21}(t_1)(X_1(t_1)-X_1(t_2))+H_{23}(t_1)(X_3(t_1)-X_3(t_2))\nonumber\\
&~+Z_2(t_1)+Z_2(t_2)/c(t_2),\label{eq:rx2}\\
Y_3(t_1)+Y_3(t_2)/c(t_2) &=
H_{33}(t_1)(X_3(t_1)+X_3(t_2))+H_{31}(t_1)(X_1(t_1)-X_1(t_2))+H_{32}(t_1)(X_2(t_1)-X_2(t_2))\nonumber\\
&~+Z_3(t_1)+Z_3(t_2)/c(t_2)\label{eq:rx3}.
\end{align}

\textit{Decoding at receiver 1}: Using the received signal at
$t_2+2$ and $t_2+3$, receiver 1 removes the interfering signals from
the other senders in (\ref{eq:rx1}). Then, we have an equation for
$X_1(t_1)+X_1(t_2)$. Using another equation for $X_1(t_1)-X_1(t_2)$
received at $t_2+1$, receiver 1 can decode both $X_1(t_1)$  and
$X_1(t_2)$.

\textit{Decoding at receiver 2}: Similarly, receiver 2 removes the interfering signals in (\ref{eq:rx2}) using the received signal at
$t_2+1$ and $t_2+3$ and decodes $X_2(t_1)$  and $X_2(t_2)$ using the received signal at $t_2+2$ and (\ref{eq:rx2}).

\textit{Decoding at receiver 3}: Similarly, receiver 2 removes the interfering signals in (\ref{eq:rx3}) using the received signal at
$t_2+1$ and $t_2+2$ and decodes $X_3(t_1)$  and $X_3(t_2)$ using the received signal at $t_2+3$ and (\ref{eq:rx3}).

According to the decoding procedure, the proposed scheme enables
each receiver to decode its 2 messages in 5 symbol times. That is,
total 6 messages are decodable over 5 symbol times and hence total
6/5 DoF is achievable.
\end{IEEEproof}

\subsection{$K$-user interference channel with delayed CSIT}\label{kudcsit}
\begin{theorem}
Total $\frac{2K}{K+2}$ DoF is achievable in a $K$-user interference
channel with delayed CSIT by ergodic IA.
\end{theorem}
\begin{IEEEproof}
For a $K$-user interference channel, two independent messages sent
at time $t_1$ and $t_2$ are decoded at each receiver over $K+2$
symbol times. As in the three-user interference channel, the time
$t_1$ and $t_2$ correspond to transmission phase 1. If the
transmitters realize the channel matrix at time $t_2$ satisfies the
condition in (\ref{eq:channel1}) and (\ref{eq:channel2}),
transmission phase 2 starts. Then, transmitter $k$,
$k\in\{1,\ldots,K\}$, sends the signal $X_k(t_1)-X_k(t_2)$ at time
$t_2+k$. Similarly to the three-user interference channel, each
receiver can decode its two messages over $K+2$ symbol times.
Therefore, total $\frac{2K}{K+2}$  DoF is achievable in a $K$-user
interference channel by the proposed ergodic IA.
\end{IEEEproof}

\begin{figure}[!t]
\vspace{0.3in}
   \centerline{\psfig{figure=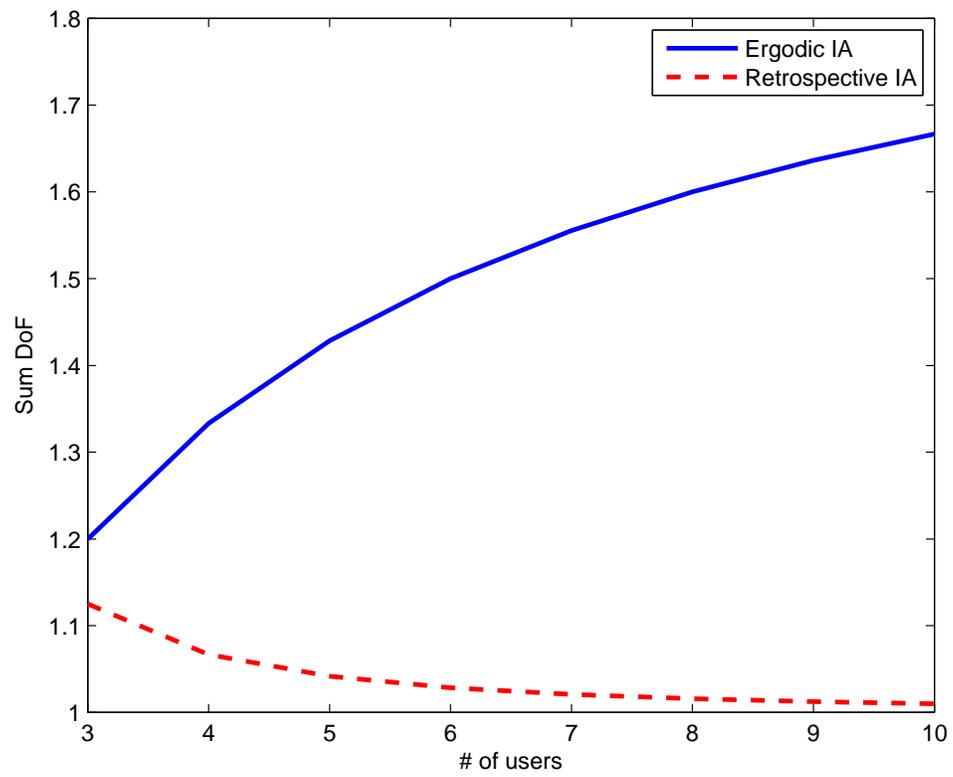,width=0.8\columnwidth} }
    \caption{Sum degrees of freedom with delayed CSIT}
    \label{Fig:P1}
\end{figure}

Fig. \ref{Fig:P1} shows the achievable DoF by the proposed ergodic
IA (solid line) and the retrospective
IA \cite{RIAK} (dashed line) with delay CSIT
according to the number of users in a $K$-user interference channel.
The total achievable DoF by the retrospective IA
is $\frac{K^2}{K^2-1}$. It starts from $\frac{9}{8}$ for a
three-user case and converges to $1$ as $K$ goes to infinity. On the
other hand, the total achievable DoF by the proposed ergodic IA
starts from $\frac{6}{5}$ and converges to $2$.

\subsection{$K$-user interference channel with delayed time index feedback}
If the receivers feed the time indices at which the condition in
(\ref{eq:channel1}) and (\ref{eq:channel2}) is satisfied, total
achievable DoF is the same as the case that delay CSIT is used.

\begin{theorem}
Total achievable DoF by the proposed ergodic IA with delay time
index feedback in a $K$-user interference channel is
$\frac{2K}{K+2}$.
\end{theorem}
\begin{IEEEproof}
The strategy in Section \ref{kudcsit} can be applied. Instead of
using the channel state information (i.e., channel matrix) to find
the time $t_1$ and $t_2$ at which the condition in
(\ref{eq:channel1}) and (\ref{eq:channel2}) is satisfied, the
receivers send the time indices $t_1$ and $t_2$ since receivers can
find them by the assumption of full CSIR. After receiving the time
indices, the transmitters realize that the previous time was $t_2$
and enter into transmission phase 2.
\end{IEEEproof}

\section{Ergodic interference alignment with delayed output feedback without CSIT}
In this section, we assume full CSIR and delayed output feedback.
The received signals themselves are fed back to the transmitters.
Each transmitter cannot use the channel information but can use only
the delayed received output feedback. It is also assumed that the
receivers can reform the output signals and feed them back to the
transmitters if necessary.

\subsection{Three-user interference channel}
Our new proposed strategy is first applied to a three-user
interference channel for better explanation of the proposed idea.

\begin{theorem}
Total $\frac{6}{5}$ DoF is achievable by the proposed ergodic IA in
a three-user interference channel when only delayed output feedback
information is available.
\end{theorem}
\begin{IEEEproof} The operation of the proposed ergodic IA is
classified into two phases:

\textit{Transmission phase 1}: The transmission phase 1 for data
transmission is continued until $t_2$. At each time until $t_2$, new
messages are continuously transmitted. For the time $t_1$ and $ t_2$
at which the channel condition in (\ref{eq:channel1}) and
(\ref{eq:channel2}) is satisfied, the received signals are given by
(\ref{rs1}) and (\ref{rs2}). Similarly to the case of delayed CIT in Section III, the transmitters cannot send the same
message at $t_1$ and $t_2$ because they cannot realize that $t_2$ is
the time instant at which the channel condition in
(\ref{eq:channel1}) and (\ref{eq:channel2}) is satisfied due to the
absence of CSIT. Therefore, the transmitters continue to send
independent messages at $t_2$. However, the receivers know that the
channel condition in (\ref{eq:channel1}) and (\ref{eq:channel2}) is
satisfied at $t_2$ owing to full CSIR so that they construct the
following output feedback information and send them back to the
transmitters after receiving the signals at $t_2$.
\begin{align}
&\mathrm{Receiver~1}:~(Y_1(t_1)+Y_1(t_2)/c(t_2))/H_{11}(t_1)\nonumber\\
&\mathrm{Receiver~2}:~(Y_2(t_1)+Y_2(t_2)/c(t_2))/H_{22}(t_1)\nonumber\\
&\mathrm{Receiver~3}:~(Y_3(t_1)+Y_3(t_2)/c(t_2))/H_{33}(t_1).\nonumber
\end{align}
Then, the transmitters can figure out that the previous time $t_2$
was the time instant at which the channel condition in
(\ref{eq:channel1}) and (\ref{eq:channel2}) is satisfied after
receiving the delayed output feedback information. However, note
that they do not know the time instant $t_1$ as well as the message
sent at $t_1$. Once after receiving the delayed output feedback
information, transmission phase 2 is entered.

\textit{Transmission phase 2}: After the output feedback signals are
received, the transmitters send their signals at time
$t_2+1,t_2+2,t_2+3$, respectively:
\begin{align}
&\mathrm{Transmitter~1~at~time}~t_2+1:~(Y_1(t_1)+Y_1(t_2)/c(t_2))/H_{11}(t_1)-2X_1(t_2)=\nonumber\\
&(X_1(t_1)-X_1(t_2))+(H_{12}(t_1)(X_2(t_1)-X_2(t_2))+H_{13}(t_1)(X_3(t_1)-X_3(t_2))+Z_1(t_1)+Z_1(t_2)/c(t_2))/H_{11}(t_1)\nonumber\\
&\mathrm{Transmitter~2~at~time}~t_2+2:~(Y_2(t_1)+Y_2(t_2)/c(t_2))/H_{22}(t_1)-2X_1(t_2)=\nonumber\\
&(X_2(t_1)-X_2(t_2))+(H_{21}(t_1)(X_1(t_1)-X_1(t_2))+H_{23}(t_1)(X_3(t_1)-X_3(t_2))+Z_2(t_1)+Z_2(t_2)/c(t_2))/H_{22}(t_1)\nonumber\\
&\mathrm{Transmitter~3~at~time}~t_2+3:~(Y_3(t_1)+Y_3(t_2)/c(t_2))/H_{33}(t_1)-2X_1(t_2)=\nonumber\\
&(X_3(t_1)-X_3(t_2))+(H_{31}(t_1)(X_1(t_1)-X_1(t_2))+H_{32}(t_1)(X_2(t_1)-X_2(t_2))+Z_3(t_1)+Z_3(t_2)/c(t_2))/H_{33}(t_1).\nonumber
\end{align}
After transmission phase 2 is completed, the transmission mode goes
back to transmission phase 1.

\textit{Decoding at receiver 1}: Linearly combining the received
signals at $t_2+1$, $t_2+2$ and $t_2+3$, receiver 1 can obtain the
values of $X_1(t_1)-X_1(t_2)$, $X_2(t_1)-X_2(t_2)$ and
$X_3(t_1)-X_3(t_2)$. By substituting the values of
$X_2(t_1)-X_2(t_2)$ and $X_3(t_1)-X_3(t_2)$ to
$Y_1(t_1)+Y_1(t_2)/c(t_2)$, the value of $X_1(t_1)+X_1(t_2)$ can
also be obtained. Then, receiver 1 can decode both $X_1(t_1)$ and
$X_1(t_2)$ because it has two independent equations on $X_1(t_1)$ and
$X_1(t_2)$ -- one is given in terms of $X_1(t_1)-X_1(t_2)$ and the
other is given in terms of $X_1(t_1)+X_1(t_2)$.

\textit{Decoding at receiver 2}: Similarly, receiver 2 can decode $X_2(t_1)$ and
$X_2(t_2)$ using linear combination and substitution.

\textit{Decoding at receiver 3}: Similarly, receiver 3 can decode $X_3(t_1)$ and
$X_3(t_2)$ using linear combination and substitution.

In this way, all 6 messages are decoded in 5 symbol times so that
total 6/5 DoF is achievable by the proposed ergodic IA.
\end{IEEEproof}

\subsection{K-user interference channel}
\begin{theorem}
When only delayed output feedback information is available at the
transmitters, total $\frac{2K}{K+2}$ DoF is achievable in a $K$-user
interference channel by the proposed ergodic IA
\end{theorem}
\begin{IEEEproof}
For a $K$-user interference channel, two independent messages sent
at time $t_1$ and $t_2$ are decoded at each receiver over $K+2$
symbol times, where the time time $t_1$ and $t_2$ correspond to
transmission phase 1. After the receivers receive the signals at
$t_2$, receiver $k$, $k\in\{1,\ldots,K\}$, feed the output
$(Y_k(t_1)-Y_k(t_2)/c(t_2))/H_{kk}(t_1)$ back to its own
transmitter. After the output feedback signals are received, the
transmitters realize that the previous time instant was $t_2$ at
which the channel condition in (\ref{eq:channel1}) and
(\ref{eq:channel2}) is satisfied and enters into transmission phase
2. In transmission phase 2, transmitter $k$ sends the signal
$(Y_k(t_1)-Y_k(t_2)/c(t_2))/H_{kk}(t_1)-2X_k(t_2)$ at only time
$t_2+k$. As in the three-user interference channel, each receiver
decode its two messages over $K+2$ symbol times. Therefore, total
$\frac{2K}{K+2}$ DoF is achievable in a $K$-user interference
channel by the proposed ergodic IA.
\end{IEEEproof}

\begin{figure}[!t]
\vspace{0.3in}
   \centerline{\psfig{figure=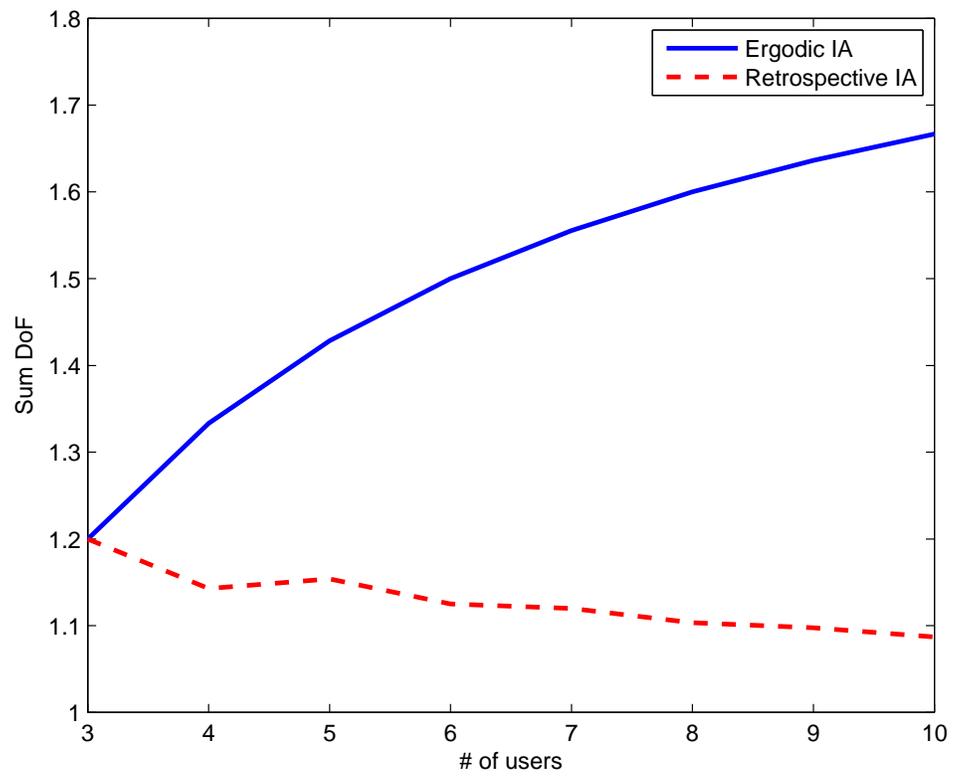,width=0.8\columnwidth} }
    \caption{Sum degrees of freedom with delayed output feedback without CSIT}
    \label{Fig:P2}
\end{figure}

Fig. \ref{Fig:P2} shows total achievable DoF by the proposed ergodic
IA (solid line) and the retrospective
IA \cite{RIAK} (dashed line) with delayed output
feedback without CSIT according to the number of users. The total
achievable DoF of the retrospective IA is
$\frac{\lceil K/2\rceil K}{\lceil K/2\rceil(K-1)+1}$. It starts from
$\frac{6}{5}$ for the three-user case and converges to $1$ as $K$
goes to infinity. On the other hand, the total achievable DoF by the
proposed ergodic IA starts from $\frac{6}{5}$ and approaches to $2$
as $K$ goes to infinity.

\section{Conclusion}
In this paper, we proposed new ergodic IA
techniques in $K$-user interference channels with delayed feedback.
Total achievable DoF by the proposed ergodic IA
is derived for two scenarios of delayed feedback -- delayed channel
information and delayed output feedback information. We showed that
total $2K/(K+2)$ DoF is achievable by the proposed schemes for both
scenarios. The proposed ergodic IA schemes
achieve higher DoF than the retrospective IA
when the feedback information is outdated.

\end{document}